\documentclass[12pt,a4paper]{article}
\usepackage{jheppub}

\usepackage{graphicx}
\usepackage{subcaption}
\usepackage{xspace}

\usepackage[utf8]{inputenc}

\usepackage[style=numeric-comp,sorting=none]{biblatex}
\abstract{
To produce the best physics results, high energy physics experiments require access to calibration and other non-event data
during event data processing.  These conditions data are 
typically stored in databases that provide versioning functionality, allowing
physicists to make improvements while simultaneously guaranteeing
the reproducibility of their results.  With the increased complexity of modern experiments, and the evolution of 
computing models that demand large scale access to conditions data, the solutions for managing this access
have evolved over time.  In this white paper we give an overview of the conditions data access problem,
present convergence on a common solution and present some considerations for the future.
}

\begin{document}

\noindent
\begin{tabular*}{\linewidth}{lc@{\extracolsep{\fill}}r@{\extracolsep{0pt}}}
 & & HSF-CWP-2017-03 \\
 & & January 16, 2019 \\ % use \date or hardwire e.g. December 15, 2017
 & & \\
\end{tabular*}
\vspace{2.0cm}

\title{HEP Software Foundation Community White Paper Working Group -- Conditions Data}

\author[1,2]{Marko Bracko,}
\author[3]{Marco Clemencic,}
\author[4,a]{Dave Dykstra,}
\author[5]{Andrea Formica,}
\author[4]{Giacomo Govi,}
%\author[6]{Hadrien Grasland,}
\author[6,b]{Michel Jouvin,}
\author[7,b]{David Lange}
\author[3,8,c]{Paul Laycock}
\author[9]{and Lynn Wood}

\affiliation[1]{University of Maribor, Ljubljana, Slovenia}
\affiliation[2]{Jožef Stefan Institute, Ljubljana, Slovenia}
\affiliation[3]{CERN, Geneva, Switzerland}
\affiliation[4]{Fermi National Accelerator Laboratory, Batavia, IL, USA}
\affiliation[5]{DSM/IRFU (Institut de Recherches sur les Lois Fondamentales de l’Univers), CEA Saclay (Commissariat à l’Énergie Atomique), Gif-sur-Yvette, France}
\affiliation[6]{LAL, Université Paris-Sud and CNRS/IN2P3, Orsay, France}
\affiliation[7]{Princeton University, Princeton, NJ, USA}
\affiliation[8]{Physics Department, Brookhaven National Laboratory, Upton, NY, USA}
\affiliation[9]{Pacific Northwest National Laboratory, Richland, Washington, USA}
\note[a]{Supported by the US-DOE, DE-AC02-07CH11359}
\note[b]{Community White Paper Editorial Board Member}
\note[c]{Corresponding Author}

\maketitle

\newpage

\section{{Introduction}}

Access to conditions data is critical to producing the best physics results from HEP experiments. 
It is somewhat surprising then that few people can accurately define what conditions data means.
The challenges for conditions
data access are many, notably the requirement to provide simultaneous read access to conditions data for distributed computing
resources at kHz rates.  This is a highly non-trivial problem that has brought difficulties, 
especially to the biggest experiments with the largest distributed computing resources; it is easy to get wrong.
Equally those shared difficulties provoked collaboration on 
shared solutions, particularly between ATLAS and CMS~\cite{Viegas:2008zz, Guida:2015gvw, 1742-6596-898-4-042047, Barberis:2015oji},
and this white paper benefits greatly from that work.

The purpose of this white paper then is to answer the following questions:
\begin{enumerate}
\item What is conditions data ?
\item What are the major use cases and challenges for conditions data access ?
\item What does best practice for conditions data access management look like ?
\item What challenges are posed by conditions data access in the future ?
\end{enumerate}

The reader should
note that this paper focuses primarily on conditions data access and therefore not all aspects of conditions data management are 
addressed\footnote{Conditions data creation workflows and management tools, authentication and authorization are all examples of topics that deserve further attention.}.

The paper is organised as follows: section~\ref{sec:CondData} answers the first 
two questions, thus defining the scope of the work presented here.  Section~\ref{sec:Arch} 
describes best practice in conditions data access management by means of
a detailed prototype. The requirements for the future success of HEP in terms of
managing conditions data access are discussed in section~\ref{sec:Future}, attempting to answer the final question in the list above,
before drawing conclusions in the final section.

\section{{Conditions data in HEP experiments}}
\label{sec:CondData}

Although conditions data is a commonly
used term in most HEP experiments, its definition remains ambiguous for many non-experts.
Such ambiguity is not helpful and has caused considerable problems in the past, some examples of which will be given later.
Conditions data is distinct from the particle collision event data collected from the experimental detectors, hereafter simply referred to as event data.
However, conditions data is not all non-event data but rather a particular subset.

From the detector expert perspective, conditions data can be broadly defined as the non-event data
required by event data-processing software to correctly reconstruct the raw detector event 
data. In the case of simulated event data, the scope extends to the correct simulation, digitisation and reconstruction of the data.  
The most obvious examples of conditions data are calibration and alignment constants, but this is not a complete picture. 
In general, from the detector perspective, conditions data is the subset of non-event data
required to maintain, operate and optimise detectors.  
This in turn can be broken down into the
following categories:

\begin{enumerate}
\def\labelenumi{\alph{enumi}.}
\item
  Detector and readout configuration parameters.
\item
  Detector Control System (DCS) data, typically monitoring values of, e.g. voltage and current that are provided directly by hardware.
\item
  Higher-level detector and system monitoring information, typically provided by custom software with the purpose of evaluating detector performance.
\item
  Detector calibration and alignment data.
\end{enumerate}

Conditions data in the detector context thus largely consists of (d), together with the subset
of (a) and (b) that are required for event data processing.  Other non-event
data may also be required for event data processing, for example particle physics accelerator
parameters.  Thus, in practice, {\it any non-event data from any source that are
required for event data processing can be considered as conditions
data.}

The use cases for conditions data
are described in section \ref{subsec:use}.
In general, conditions data vary with time but with a granularity
much coarser than a particle collision event, typically having an interval of validity ranging from one year to one data acquisition 
run\footnote{The minimum granularity is of course arbitrary, but care should be taken to keep the conditions data volume reasonable, further discussed in section \ref{subsec:volume}.}.

\subsection{Workflows and use cases}
\label{subsec:use}

The original use cases for conditions data
were in offline software, but with the advent of software-based high level triggers this is no longer
the only use case.  
Broadly speaking, the major conditions data access use cases are:

\begin{enumerate}
\def\labelenumi{\arabic{enumi})}
\item
  Online event data processing of raw detector event data
\item
  Offline event data processing of 
\begin{enumerate}
  	\item raw detector event data
	\item fully reconstructed event data
	\item simulated event data
\end{enumerate}
\item
  Higher-level analysis of processed event data
\end{enumerate}

To a large extent this paper deals with the first two use cases while
analysis is covered briefly in section \ref{subsec:analysis}.
Online event data processing typically has infrequent updates of conditions,
with stability and predictability preferred over optimal performance. The exceptions are those conditions that are critical to the
performance of trigger algorithms, e.g. the position of the beam-spot in the case of ATLAS and CMS.  Online event data processing runs on dedicated 
resources with a dedicated server for conditions data access.
%By definition, online event data reconstruction also has no requirement for versioning of conditions,
%there is only one version of the conditions that were used for triggering
%an event.  In practice versioning of conditions data is used
%by the online conditions experts in order to compare and validate different versions of conditions
%\footnote{Reprocessing of event data using the software trigger is not treated as an online use case in this context.}.

Offline event data processing in general starts with the reconstruction of raw detector event data, while for simulated data it requires
simulation, digitisation and reconstruction of the generated events.  In modern experiments,
there is usually subsequent data reduction of the output of fully reconstructed event data to produce smaller samples that are 
better targeted for physics analysis.
Offline event data processing is typically repeated several times following improvements in software and/or conditions.
It generally involves distributed event data processing on the grid, making it the most challenging use case for conditions data access, requiring 
conditions data caching to run at scale.
HPC and similar off-grid resources are also viable offline event data processing locations and have their own special requirements,
in particular a lack of external connectivity.

%A non-exhaustive list of important workflows follows:
%\begin{enumerate}
%\def\labelenumi{\arabic{enumi})}
%\item
%  Subsystem calibration: conditions data creation and testing (including
%  the ability to access a copy of the conditions stored locally);
%  uploading conditions to the production database.
%\item
%  Prompt event data processing, with conditions updates adhering to strict
%  protocols:

%  \begin{enumerate}
  %\def\labelenumii{\alph{enumii}.}
%  \item
%    the software trigger: running on dedicated resources with a
%    dedicated server.
%  \item
%    similarly for prompt offline event data processing, including the
%    calibration determination.
%  \end{enumerate}
%\item
%  Offline event data processing, including Monte Carlo simulation, using
%  pre-determined conditions.

%  \begin{enumerate}
  %\def\labelenumii{\alph{enumii}.}
  %\setcounter{enumii}{2}
%  \item
 %   Importantly this generally involves distributed event data processing on
 %   the grid where caching is required to be able to run at scale.
%  \item
%    HPC and similar off-grid resources are also viable event data
%    processing locations and have their own special requirements,
%    in particular a lack of external connectivity.
%  \end{enumerate}
%\end{enumerate}

\subsubsection{Conditions for analysis}
\label{subsec:analysis}

Versions of conditions data for analysis have often been managed in ad hoc ways, if at all, 
with hard-coding of conditions data into analysis software releases being considered 
normal\footnote{The situation is improving for running experiments, albeit with different solutions than those used for offline event data processing.}.  
Belle~II~\cite{Ritter:2018jxh, Wood:2017szx} has adopted 
a different approach where conditions data for analysis will be treated in the same way as the rest of their 
conditions data.  They will use
a conditions data access management system similar to the one presented here and, given the growing interest in the analysis community,
feedback on the suitability of these solutions is expected in the near future.

\subsection{Data volumes, read and write rates}
\label{subsec:volume}

Typically, write-rates for conditions data must support of the order of 1
Hz to ensure good support for several independent systems writing
conditions data every minute, with the majority of conditions data
being updated much less frequently than this. On the other hand,
read-rates up to several kHz must be supported for distributed computing
workflows, where thousands of jobs needing the same conditions data may
start at the same time. Conditions data are typically written once
and read frequently.

Conditions data tend to scale in a controlled way during the lifetime of
an experiment, typically producing data volumes of gigabytes to terabytes. 
It is worth noting however that for ATLAS, where the same database
instances were used for conditions data as well as the DCS and trigger
data, the DCS and trigger data dominated the offline and online database
instances, respectively. Raw DCS data, which typically have very high granularity, 
are not generally needed for event data processing.  Smoothed DCS data which have been 
treated to provide a granularity appropriate for event data processing are used instead.  
High granularity raw DCS data are vital for understanding problems with detectors, but
mixing these two very different use cases in the same database schema led to several difficulties. 
Firstly it was difficult to optimise the database tables, and secondly conditions data access for offline event data processing 
required overly complicated database clients and infrastructure to cope with offline conditions 
data access at scale.  It is therefore strongly recommended by this
working group to factorise conditions data access from other use cases, even
where these are related.

\section{{Conditions Database Archetype}}
\label{sec:Arch}

The archetypal solution to a conditions database management system
(CDMS) is shown below. The conditions data payloads are stored in a
master database and are accessed using a client-server design through a
REST interface. All of the experiments consulted gave feedback that achieving a
high degree of separation between client and server was very desirable.
Due to the read-rate requirements, caching is extremely important and
good experience was seen when using web-proxy caches, e.g.~the Squid
cache shown here. Some key design principles are detailed in the
rest of this section.

\begin{figure}[h]
\begin{center}
\includegraphics[width=0.8\textwidth]{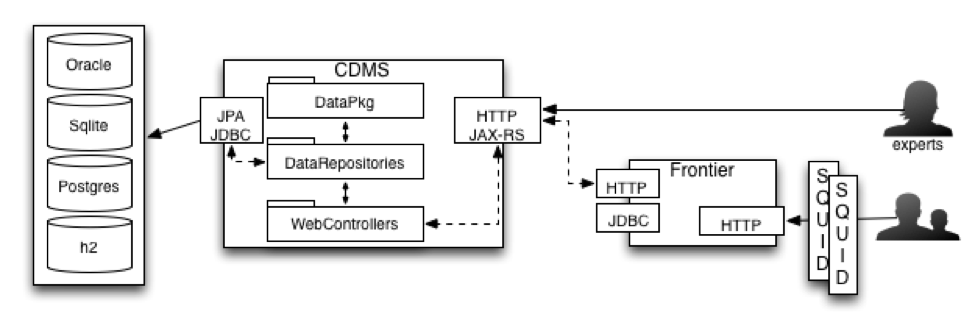}
\caption{The Conditions Database Management System archetype.}
\label{fig:meta}
\end{center}
\end{figure}

\subsection{Payload technology}

Experiments will inevitably choose their favourite payload technology.
This working group recommends placing most emphasis on
homogeneity and long-term maintenance when making this choice.
Inhomogeneity and home-grown solutions all place additional burdens on
projects that typically lack the resources to support this after the
initial build and commissioning phase of an experiment. CMS~\cite{Guida:2015gvw} has very
good experience of removing choice and only supporting boost-serialised
C\texttt{++} objects, with all classes belonging to one package in their software
framework. Such a strategy lends itself more readily to long-term
maintenance and minimises hurdles to data preservation. It is noted that
there are payload formats used routinely in industry which would lend
themselves more to higher-level functionality without the need of the
software framework but, while this is attractive, the choice of format
tends to be driven by software framework developers.

One very important consideration for payload technologies is the issue of
schema evolution.  This is further exacerbated in the (frequent) case that the 
experiment uses the same framework for the online and offline event data processing use cases.  
Conditions payloads are determined offline and therefore their versioning
follows the offline release which will usually evolve more rapidly than the online
release.  This implies a forward compatibility constraint such that the online software
can correctly process conditions payloads generated by the offline software which may use a
later version.  In the case of CMS, as boost does not support forward-compatibility
in its schema evolution, conditions payloads for a particular set of event data must be written by a
version of the software framework that is at most as old as the version used by the online software 
when recording those event data.

\subsection{Database back-end}

One of the key features of the design is that it is agnostic to
particular choices of database back-end. This was also one of the key
features of COOL~\cite{Valassi:2008zka, LCGCOOL:Homepage}, but due to the lack of caching in COOL, the queries
themselves had to become more complicated. Thus on ATLAS, which uses an
Oracle back-end, a significant amount of Oracle DBA effort was required
to tune the queries and make them performant. It is therefore important
to realise that real flexibility with respect to choices in database
back-ends only comes when the system as a whole is simplified.

\subsection{Client-side requirements}

The client layer should be as simple as possible and should be as
agnostic of the rest of the architecture as it is possible to be in
order to improve maintainability. The database insertion tools in
particular benefit from adopting a simple, e.g. REST, interface. The
client layer needs to take care of payload deserialisation, as the
remaining architectural components will deal with serialised objects.
Experience also shows that clients should be able to manage multiple
proxies and servers to provide robustness against server failures.

\subsection{Caching}

CMS and ATLAS absolutely require an intermediate layer, Frontier~\cite{Dykstra:2011zz, Frontier}, between the client and
server to provide caching capabilities. Considering offline event data processing using distributed computing resources, 
where thousands of jobs start at the same time and require the
same conditions data, this is a clear requirement and one that can be
well met using web-proxy technologies. An alternative solution would be
to use a distributed file-system with good caching capabilities, and the
LHCb~\cite{ClemencicCHEP2018} and ALICE~\cite{ALICE-CDB} experiments have gained experience using CVMFS~\cite{1742-6596-331-4-042003}. 
The simplicity of this solution makes it very attractive and it has thus been adopted as a
strategy by the NA62~\cite{LaycockCHEP2018} experiment. This in turn suggests that, as a design
requirement, it should be possible to represent a conditions database on
a file-system. The primary challenge here is to make the file-system mapping
use the CVMFS caching layers efficiently. Several experiments also have
experience using SQLite replicas, including for HPC usage. These are attractive for 
workflows where the exact subset of conditions is known in advance, e.g. some simulated data
workflows, but in general a performant caching layer is preferable thanks to its flexibility.

\subsection{Data Model}

The data model for conditions data access management is an area where the
experiments have almost universally agreed on a design, shown in figure \ref{fig:meta}. 
\begin{figure}[h]
\begin{center}
\includegraphics[width=0.8\textwidth]{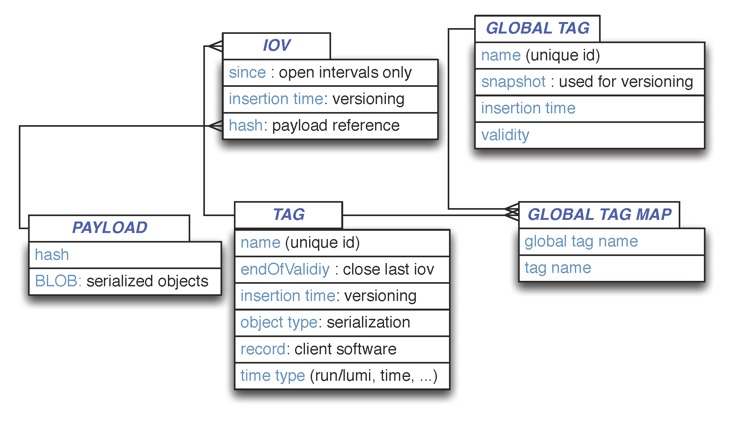}
\caption{The data model for conditions data access management
%. The point of entry is specified by a global tag, which
%for a particular type of conditions data payload and a particular interval of validity, maps to one conditions data payload via a cacheable hash 
(see text).}
\label{fig:meta}
\end{center}
\end{figure}

A global tag is the top-level
configuration of all conditions data. For a given payload 
type\footnote{A payload type can be e.g. the muon detector alignment.} 
and a given
interval of validity\footnote{A period of time for which a conditions data payload is considered valid.}, a global tag will resolve to one, and only one,
conditions data payload. The \textbf{Global Tag} resolves to a
particular payload type \textbf{Tag} via the \textbf{Global Tag Map} table. A
payload type \textbf{Tag} consists of many non-overlapping intervals of validity or entries in
the \textbf{IOV} table. Finally, each entry in the \textbf{IOV} table
maps to a payload via its unique hash key in the \textbf{Payload} table.
A relational database is a good choice for this design.

The data model design has several key features. Firstly, conditions data payloads
are uniquely identified by a hash which is the sole reference to any
given conditions data payload. The payload data has been separated from
the data access management metadata and can both in principle and in practice be placed in a
separate storage system. This could also be important for data
preservation, as the entire metadata component will occupy a trivial
data volume and could exist in, e.g. an SQLite file, while the payload
storage could be handled separately. Secondly, IOVs are resolved
independently of payloads and are also cacheable. Efficient
caching is a key design requirement for any conditions database access system that
must support high rate data access.

\subsection{A git-based approach}

For workflows which are completely offline and asynchronous with respect
to data-taking, a subset of the problems discussed previously,
LHCb has adopted a different approach. Using git as the
versioning system, conditions data payloads are placed in a directory structure, one
directory for each payload type.  A file is used to map timestamps to payload
files, and a simple format is used to allow a level of indirection to
improve performance. The file format allows a timestamp to point either
directly to a payload file or to a directory, thus allowing partitioning
of the lookup. Versioning is  taken care of by creating a git tag,
which is equivalent to a global tag, while the conditions data payloads
are stored on CVMFS.

\section{{Future and roadmap}}
\label{sec:Future}

The conditions data access management model described here will be tested by 
Belle~II in 2019 and by CMS and ATLAS in LHC Run 3.  Assuming they work as expected, 
this will meet the conditions data access 
performance demands of HEP for the coming decade.  Even at the HL-LHC~\cite{HL-LHC}, conditions data volumes are
not expected to exceed a Terabyte per year and the rate of requests,
determined by the computing resources of the experiments, are expected to peak at tens of kHz.  
Based on experience with similar conditions data access management systems, particularly the Run 2 CMS implementation, the
outlook is positive.

Nevertheless there are concerns.
The most important issue for experiments will be maintenance and
operation in the face of evolving hardware and infrastructure and a dwindling number of conditions data management experts.
These concerns provide even more motivation to consolidate the conditions data access model to a simple and modular design, such as that
presented here. Going further, the logical next step of the collaborative, cross-experiment work started by members of this working group would ultimately 
result in an experiment-agnostic HEP conditions data service.  The challenge for the community is realising that ambition.

\section{{Conclusion}}

Conditions data management is an important component of HEP software and one that
has relatively few experts.
While conditions data volumes are easily accommodated by several database
technologies, conditions data access use cases can be demanding. In particular, 
it is critical that solutions support access rates at the level of tens of kHz.

Several experiments have converged on a common design for conditions data access management.  
A key feature is a high degree of separation between a relatively simple client and the server, 
a design that is well-suited to REST interfaces.  Loosely-coupled industry standard components provide much of the robust software stack.  An intermediate layer with caching capability
is required to support kHz rates of read requests, and web-proxies have performed
very well. Combined
with a relational data model that isolates payloads from metadata, the design can support a
wide variety of HEP workflows both now and in the coming decade.  The
main challenges the community faces will be finding resources for the maintenance and
operation of this solution in the face of evolving hardware and infrastructure.
Having achieved a common design, the solution to this problem may be the logical next step, a common HEP conditions data service.

\printbibliography[title={References},heading=bibintoc]

\end{document}